\documentclass[twocolumn,english,prb]{revtex4-2}
\usepackage{amsmath}
\usepackage{amssymb}
\usepackage{graphicx}

\makeatletter
\usepackage{braket}

\makeatother

\usepackage{babel}
\begin{document}
\title{Counterdiabatic route for preparation of state with long-range topological
order}
\author{Sanjeev Kumar}
\affiliation{Department of Theoretical Physics, Tata Institute of Fundamental Research,
Homi Bhabha Road, Navy Nagar, Mumbai 400005, India}
\author{Shekhar Sharma}
\affiliation{Department of Theoretical Physics, Tata Institute of Fundamental Research,
Homi Bhabha Road, Navy Nagar, Mumbai 400005, India}
\author{Vikram Tripathi}
\affiliation{Department of Theoretical Physics, Tata Institute of Fundamental Research,
Homi Bhabha Road, Navy Nagar, Mumbai 400005, India}
\date{\today}
\begin{abstract}
We propose here a counterdiabatic (CD) strategy for fast preparation
of a state with long-range topological order by magnetic field tuning
of an initial separable state. For concreteness, we consider the ground
state of the honeycomb Kitaev model whose long-range topological order
together with the anyonic excitations make it an interesting candidate
for fault-tolerant universal quantum computation and storage. The
required CD perturbation is found to be local, having the form of
the off-diagonal exchange interactions reminiscent of trigonal deformations
in Kitaev Hamiltonians. We show that the counterdiabatically produced
state can have high fidelity and retain numerous desired entanglement
properties. 
\end{abstract}
\maketitle

\section{Introduction}

The usefulness of a quantum computer depends on the ability to exploit
the quantum entanglement and linear superposition absent in their
classical counterparts. How well the entanglement property of a quantum
many body state can be utilised in applications depends on purity
of the state, details of entanglement \cite{kitaev2003fault,Gross_too_entangled}
and how feasibly it can be prepared in a short time since long duration
preparation protocols may lead to environment induced decoherence
\citep{TwoatomEntaglement_prep_nature,ChernInsulator_prep_prb,adiab_topo_prep_prl,decoherence_cd_iop,STA_concepts}.
Ideally one should be able to prepare such states from an easily accessible
initial quantum state through adiabatic tuning of a suitable Hamiltonian.
The quantum adiabatic theorem is a no-go theorem for fast protocols
as they would result in non-adiabatic excitations. In contrast to
the adiabatic evolution of Hamiltonian, shortcut to adiabaticity (STA)
or counterdiabatic (CD) are transitionless driving protocols which
provide the means to ramp up the evolution without compensating the
purity of the desired final state \citep{STA_concepts,Polkovnikov_lectures,SHM_STA_prl,Jarzynski_prx}.
CD driving suppresses these non-adiabatic excitations by adding an
auxiliary field, $\hat{H}_{1}(t)$ to the system Hamiltonian, $\hat{H}_{0}(t)$.
With this auxiliary field, even for a very rapid protocol (as compared
to its adiabatic counterpart), the system always traverses the adiabatic
manifold of $\hat{H}_{0}(t)$ and certainly not of $\hat{H}_{0}(t)+\hat{H}_{1}(t)$.
The explicit expression for the counterdiabatic perturbation \citep{Jarzynski_pra,Berry_iop,CDBO_STA_iop}
is given by
\begin{equation}
\hat{H}_{1}(t)=i\hbar\sum_{m\neq n}\frac{\Ket{m}\Bra{m}\partial_{t}\hat{H}_{0}(t)\Ket{n}\Bra{n}}{E_{n}-E_{m}},\label{eq:H1}
\end{equation}
where $\Ket{m}$ denotes the instantaneous eigenstate of $\hat{H}_{0}(t)$
with eigenvalue $E_{m}$. Physically, the CD assistance does not work
through increasing the spectral gap (thereby reducing Landau-Zener
transitions) but through suppression of matrix elements that would
connect the states in the adiabatic manifold to those outside. The
expression Eq. (\ref{eq:H1}) is reminiscent of the Berry curvature.
Indeed, the transitionless CD driving compensates for the Berry curvature
\citep{Hartmann_iop} resulting in higher fidelity at the end of the
protocol. The entire spectrum of $\hat{H}_{0}(t)$ is required to
contruct $\hat{H}_{1}(t)$. Moreover, the above expansion suffers
from the issue of vanishing denominators in quantum many-body chaotic
systems \citep{Polkovnikov_lectures}. Non-locality of the CD term
and exponential sensitivity to any perturbation in the many-body Hamiltonian
\citep{Polkovnikov_waiter_pnas} is a consequence of constraining
the large number of degrees of freedom of the system to the transitionless
manifold. This limits the applicability of fast protocols to small
few-level systems and the thermodynamic limit is out of question \citep{ninesite_cd_prr}.
The perturbation $\hat{H}_{1}(t)$ suppress excitations for all $\Ket{m}$'s
and not just for some special state, for e.g., well separated ground
state of the system.

As was pointed out in Ref. [\onlinecite{Polkovnikov_waiter_pnas}], restriction
of exponentially large degrees of freedom in many-body dynamics is
not always the goal. Practically, the exact and formal rigidness of
Eq. (\ref{eq:H1}) is relaxed by focusing on a specific state only
and considering some local operators as an approximation for transitionless
driving. Here we focus on the fast preparation of the ground state
of the Kitaev Hamiltonian using the CD protocol. The Kitaev model
is an integrable 2D system of spin-1/2 particles on the honeycomb
lattice interacting with peculiar Ising-like direction dependent local
interactions \citep{kitaev2006anyons}. The model exhibits the spin
fractionalization phenomenon with no magnetic order, and elementary
excitations consisting of free Majorana fermions and \emph{gapped},
quantized half-vortices (visons). Depending on the interaction parameters,
the half-vortices are Abelian or non-Abelian anyons \citep{kitaev2006anyons}.
For isotropic Kitaev interactions, the Majorana fermions are linearly
dispersing and massless, but they become massive above a small characteristic
magnetic field. The ground state has long range topological order
\citep{kitaev2006anyons,Kitaev_Preskill}, signified by a finite topological
entanglement entropy $\gamma=\ln2,$ which is not destroyed by small
magnetic fields. Owing to the non-local entanglement and long-range
topological order, both types of anyons are useful resources for universal
quantum computation, providing fault-tolerant quantum memory and quantum
gates realisation: the Abelian ones are likely to be easily accessible
in experiments \citep{lloyd2002quantum} while the non-Abelian kind
is more useful (although less readily realizable) for the quantum
computation purposes \citep{kitaev2006anyons}.

Adiabatic preparation of highly entangled quantum states are typically
hard to implement owing to the problem of exponentially small (in
system size $N$) spectral gaps near the ground state (see e.g. Ref. [\onlinecite{altshuler_anderson_pnas}]).
However, the excitation gap near the ground state of the isotropic
Kitaev model vanishes only as $1/\text{poly}[N],$ and, upon the introduction
of a magnetic field above a certain small threshold, a finite (Majorana)
gap separates its ground state from the rest of the spectrum \citep{kitaev2006anyons,magnetic_Kitaev_prb},
while the vison gap is essentially unchanged. This will be our regime
of interest. Since the vison excitations tend to destroy the long
range topological order, the finite vison gap is a desirable feature.

We present a local CD protocol for high fidelity preparation of the
Kitaev ground state, on timescales significantly shorter than that
permitted by the quantum adiabatic theorem. The topological entanglement
entropy, $\gamma,$ and the (half-vortex) plaquette fluxes $W$ at
the end of the protocol are found to be much closer to the equilibrium
values compared to that obtained in the same time without the CD protocol.
The CD interactions in our model have the form of certain off-diagonal
exchange interactions \citep{trigonal1,trigonal2,trigonal3} in Kitaev
systems commonly associated with trigonal deformations. Besides their
physical realisation in Kitaev materials, such interactions in addition
to the Kitaev coupling can be implemented using superconducting quantum
circuits \citep{quantum_simulation_PhysRevA.99.012333,quantum_simulation_PhysRevB.81.014505}.

The rest of the paper is organised as follows. In Sec. \ref{sec:Model}
we introduce our model Hamiltonian and CD protocol. Section \ref{sec:Results}
presents our calculations of the state fidelity and other properties
such as the topological entanglement entropy and the plaquette flux
expectation value. This highlights the difference between our protocol
and the naive unassisted protocol in the results section. Section
\ref{sec:Discussion} summarizes our findings and contains a discussion.

\section{Model and counterdiabatic perturbation \label{sec:Model}}

\begin{figure}
\includegraphics[width=0.9\columnwidth]{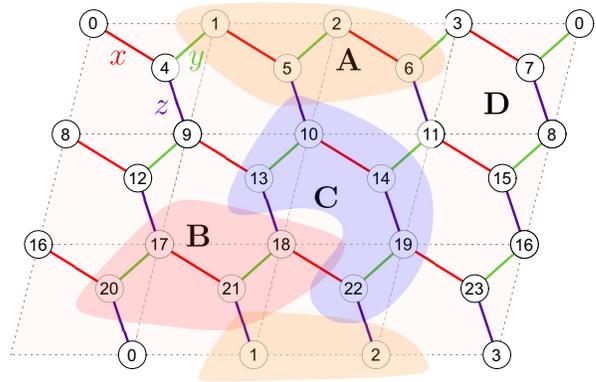}\caption{\label{fig:lattice}Schematic of the honeycomb lattice with nearest-neighbour
Kitaev interactions described in Eq. \ref{eq:Hk}. Exact diagonalization
calculations have been performed with a 24-site cluster with toroidal
boundary conditions, and benchmarked with density matrix renormalization
group calculations on larger sizes (see Appendix \ref{sec:AppendixB}). For computation
of topological entanglement entropy using the Preskill-Kitaev construction (see Sec. \ref{sec:Results}), partition
of cluster into subsystems $A=\{1,2,5,6\}$, $B=\{17,18,20,21\}$,
$C=\{10,13,14,19,22\}$ and $D$ is shown where $D$ encloses the
lattice sites not in $A,$ $B$ and $C.$}
\end{figure}

In our study, we consider the following Hamiltonian, 
\begin{equation}
\hat{H}_{0}(\lambda)=\hat{H}_{\text{k}}+(\lambda+\delta)\hat{H}_{\text{m}},\label{eq:H0}
\end{equation}
where $\hat{H}_{\text{k}}$ the Kitaev Hamiltonian on the honeycomb
lattice given by
\begin{equation}
\hat{H}_{\text{k}}=-J\sum_{\left\langle ij\right\rangle _{\gamma-\text{{link}}}}\sigma_{i}^{(\gamma)}\sigma_{j}^{(\gamma)},\label{eq:Hk}
\end{equation}
Here $i$ labels the sites and $\left\langle ij\right\rangle _{\gamma-\text{{link}}}$
denotes the nearest neighbours $i,j$ on a link $\gamma=x,y,z$ as
shown in Fig. (\ref{fig:lattice}). $\hat{H}_{\text{m}}$ corresponds
to an external Zeeman field, 
\begin{equation}
\hat{H}_{\text{m}}=B\sum_{i}\left(\sigma_{i}^{x}+\sigma_{i}^{y}+\sigma_{i}^{z}\right),\label{eq:Hm}
\end{equation}
and $\lambda(t)\in[0,1]$ parametrizes the protocol for the Hamiltonian
evolution. 
 We set the Kitaev interaction scale $J=1.$ The system is initiated
in a high external magnetic field, $B\gg1,$ such that the ground
state is a product state, and gapped. The time dependent part, $\lambda(t)$
is a smooth yet fast ramp evolving from $1$ to $0$ in a time $\tau,$
and $\delta$ is a small positive constant such that the magnetic
field at the end of our protocol is finite, but small, $B\delta\ll1$.
The spectrum remains gapped throughout the parametric evolution (see
below), decreasing monotonously as $\lambda$ decreases. 

For transitionless evolution, we now introduce the CD perturbation
to $\hat{H}_{0}.$ The CD term of Eq. (\ref{eq:H1}) is expressed
through a gauge potential, $\hat{A}_{\lambda},$ such that $\hat{H}_{1}(\lambda)=\dot{\lambda}\hat{A}_{\lambda}:$
\begin{equation}
\hat{H}_{\text{CD}}=\hat{H}_{0}+\dot{\lambda}\hat{A}_{\lambda}.\label{eq:Hcd}
\end{equation}
We require the prefactor $\dot{\lambda}$ to vanish at the end points
of the protocol; this condition ensures we begin and end in the ground
state manifold of $\hat{H}_{0}.$ The gauge potential can be expressed
as a sum of nested commutators \citep{Polkovnikov_floquet_cd_prl},
\begin{equation}
\hat{A}_{\lambda}=i\sum_{k=1}^{\infty}\alpha_{k}\underbrace{[\hat{H}_{0},\hat{H}_{0},...[\hat{H}_{0}}_{2k-1},\partial_{\lambda}\hat{H}_{0}]]],\label{eq:A}
\end{equation}
where we have suppressed $\hbar.$ The above series expansion gives
the exact gauge potential for a gapped system, and higher order commutators
generate increasingly nonlocal contributions. As an approximation,
Eq. (\ref{eq:A}) is truncated after a certain order to ensure local
interactions, while still suppressing excitations to give a reasonable
fidelity of a quantum state, which in this paper is set at $\geq0.5.$
The variational parameters, $\alpha_{k}$'s are then found by minimizing
$S=\bigl\langle\hat{G}^{2}\bigr\rangle-\bigl\langle\hat{G}\bigr\rangle^{2}$,
where 
\begin{equation}
\hat{G}=\partial_{\lambda}\hat{H}_{0}-i[\hat{H}_{0},\hat{A}_{\lambda}],\label{eq:G}
\end{equation}
and $\left\langle \text{ }\right\rangle $ denotes averaging with
respect to the Boltzmann weight, $\exp(-\beta\hat{H}_{0})$. The minimization
condition ensures transitions due to non-zero off-diagonal elements
in the instantaneous Hamiltonian are suppressed \citep{Polkovnikov_lectures}.
For ease of calculation we focus on the infinite temperature limit
($\beta\rightarrow0$), where the problem reduces to minimizing $S=\text{Tr}\bigl[\hat{G}^{2}\bigr]$.
Note the infinite temperature is not ideal for ground state preparation
as it treats the excited states on the same footing. We show the CD
assistance produces desirable results even in this worst scenario
limit. It has been shown that for the one dimensional Kitaev model,
the CD Hamiltonian is of $M-$body interaction type thus limiting
its practicality \citep{kyaw_iop} for cluster state generation. Limiting
ourselves to two-body interactions only, we retain in Eq. (\ref{eq:A})
only the leading term and obtain \begin{widetext}
\begin{figure*}
\includegraphics[width=1\textwidth]{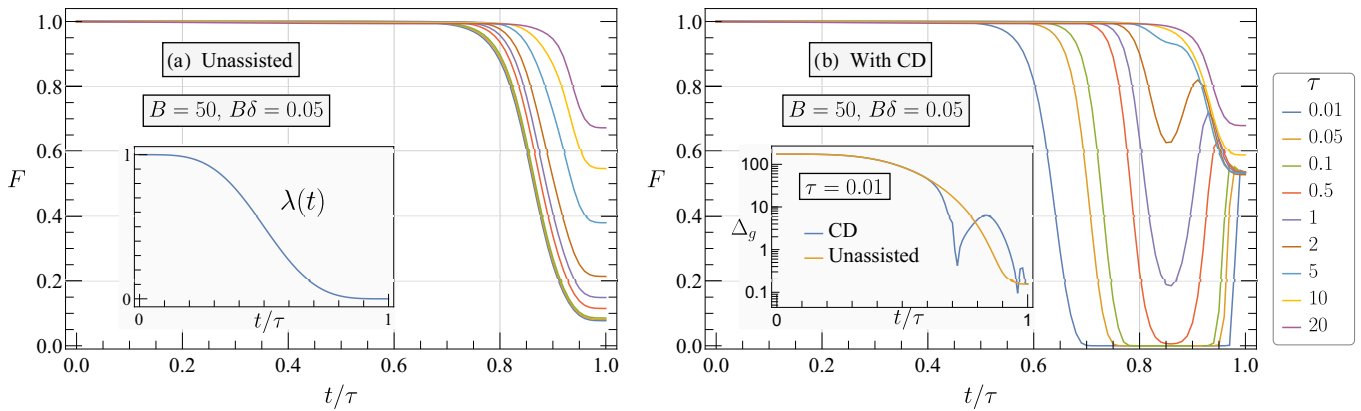}\caption{\label{fig:F}Plots showing time evolution of the fidelity, $F(t),$
for various duration-of-protocol values, $\tau$. In (a) we show the
results for the unassisted protocol i.e., parametric evolution of
the Hamiltonian without the CD term, while (b) shows results using
the CD assisted protocol. In both cases, we choose an initial large
Zeeman field, $B=50,$ which decreases to a small value, $\delta B=0.05,$
at the end of the protocol. Durations varying from $\tau=0.01$ to
$\tau=20$ are shown, spanning both sides of the validity condition for the adiabatic theorem.
For smaller values of $\tau,$ CD greatly aids in suppressing the
transitions to excited states, resulting in much larger fidelities
compared to the unassisted case. For $\tau=20$, it can be seen that
there is no significant difference in the fidelities obtained. The
inset in (a) shows the dimensionless smooth ramp $\lambda(t)=\cos^{2}\left(\frac{\pi}{2}\sin^{2}\left(\frac{\pi t}{2\tau}\right)\right)$
used in our calculations. Inset in (b) shows the energy gap between
the ground state and the first excited state in the two protocols
as a function of time $t.$ Clearly, CD does not assist in reducing
the minimum energy gap.}
\end{figure*}
\begin{equation}
\hat{A}_{\lambda}^{(1)}=\frac{B/J}{18(\lambda+\delta)^{2}(B/J)^{2}+10}\left\{ \sum_{\left\langle ij\right\rangle _{x-\text{{link}}}}\left(\sigma_{i}^{x}\sigma_{j}^{y}-\sigma_{i}^{x}\sigma_{j}^{z}\right)+\sum_{\left\langle ij\right\rangle _{y-\text{{link}}}}\left(\sigma_{i}^{y}\sigma_{j}^{z}-\sigma_{i}^{y}\sigma_{j}^{x}\right)+\sum_{\left\langle ij\right\rangle _{z-\text{{link}}}}\left(\sigma_{i}^{z}\sigma_{j}^{y}-\sigma_{i}^{z}\sigma_{j}^{x}\right)+i\leftrightarrow j\right\} .\label{eq:A1}
\end{equation}
\end{widetext} 

The above gauge potential, $\hat{A}_{\lambda}^{(1)}$ resembles the
$\Gamma^{'}$-interaction in Kitaev systems with the difference that
Eq. (\ref{eq:A1}) has asymmetric terms while $\Gamma'$-interaction
has symmetric ones \citep{trigonal1,trigonal2,trigonal3}, and associated
with trigonal distortions in the lattice. Using Eq. (\ref{eq:A1})
in Eq. (\ref{eq:Hcd}) gives the CD Hamiltonian which we use in the
rest of the paper. Numerical calculations are performed by exact diagonalisation
of a 24-site cluster (see Fig. (\ref{fig:lattice})) using Quspin
\citep{quspin2017,quspin2019}. For benchmarking, we compare the energy
gap in our system with that of a larger 144-site cluster (obtained
via DMRG, see Appendix \ref{sec:AppendixB}), and find that the two are in agreement. We
thus conclude that the system is gapped throughout the range of magnetic
fields we study - this is in contrast with some calculations \citep{Nandini_pnas}
in the recent literature (based on apparently power-law decay of ground
state spin correlators obtained using DMRG) that claim the existence
of a gapless phase in the range $0.2\lesssim B\lesssim0.3.$ The finite
spectral gap even in the thermodynamic limit is of relevance to our
problem since a vanishing gap at intermediate magnetic fields would
invalidate the counterdiabatic protocol.

\section{Results \label{sec:Results}}

\begin{figure*}
\includegraphics[width=1\textwidth]{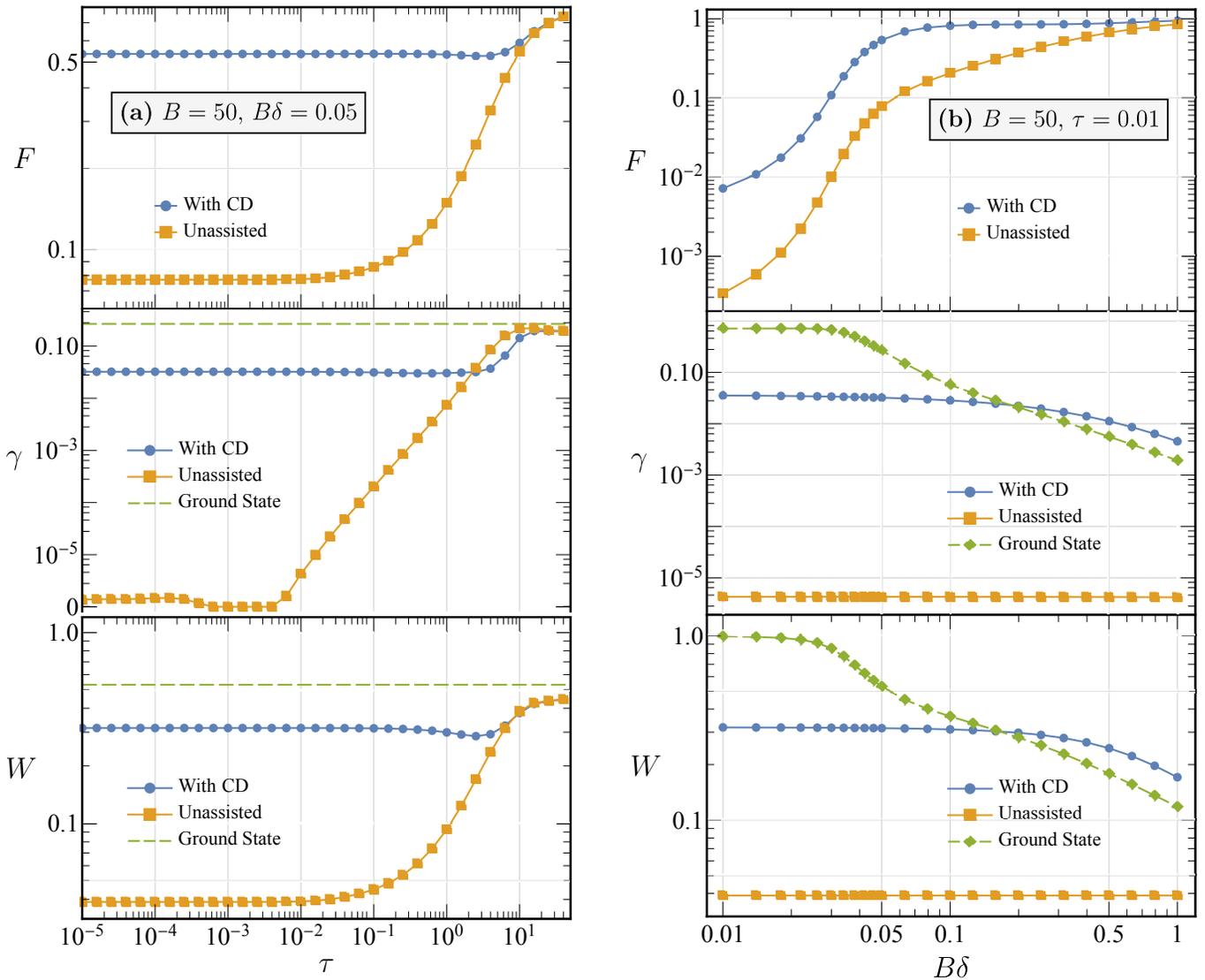}\caption{\label{fig:FgammaW}Plots showing the fidelity ($F$), entanglement entropy
($\gamma$) and expectation value of the $\hat{W}$ (plaquette flux) operator, ($W$)
calculated for the quantum state obtained via CD assisted and naive,
unassisted protocol. In (a) these quantities are shown as a function of a wide range of protocol time
duration, $\tau$ spanning both sides of the validity condition of the adiabatic theorem. 
Dashed line shows the pure ground state
value corresponding to $\delta\hat{H}_{\text{m}}+\hat{H}_{\text{k}}$.
Unassisted protocol yield close-to zero values for $\gamma$ and $W$
owing to lesser fidelity as compared with the CD assisted protocol. $\gamma$ values smaller than $10^{-6}$ are suppressed to zero by switching from log scale to linear scale.
(b) Show the same quantities as a variartion of final magnetic field
in the system $B\delta.$ The quantities measure start converging
for $\tau=10.$}
\end{figure*}

Below we show the numerical results for the overlap of the time evolved
state, $\Ket{\psi(\lambda(t))}=U(t)\Ket{\psi(\lambda(0))},$ with
the ground state $\Ket{\phi_{\text{GS}}(\lambda(t))}$ of the instantaneous
Hamiltonian. The fidelity, $F,$ is defined as $F=\left|\Braket{\psi|\phi_{\text{GS}}}\right|^{2}$.
A smooth ramp with vanishing time-derivative at the end points ensures
the initial and final Hamiltonians are the same in the unassisted
and CD protocols. For concreteness we choose $\lambda(t)=\cos^{2}\left(\frac{\pi}{2}\sin^{2}\left(\frac{\pi t}{2\tau}\right)\right)$
for $t\in[0,\tau]$ as shown in the inset of Fig. \ref{fig:F}a. Figure
\ref{fig:F} shows the fidelity of the evolving quantum state in the
(a) unassisted and (b) CD assisted protocols for various time-durations,
$\tau.$ We note that for $\tau\ll1,$ the fidelity in the CD assisted
protocol falls sharply around $t\gtrsim\tau/2$ (even falling to zero
for the shorter durations) before jumping to values significantly
larger than the unassisted case towards the end of the protocol. The
vanishing fidelity during intermediate times is not on account of
any closure of the spectral gap (see inset of Fig. \ref{fig:F}b)
but rather shows that the time evolved state in the CD protocol overlaps
poorly with the ground state of the instantaneous Hamiltonian for
these intermediate times. However, for $\tau\gg1$, the two protocols
do not show a significant difference. This is in accordance with the
fact that for slow variation of the system parameters, the CD Hamiltonian
approaches the adiabatically varying Hamiltonian. From Fig. \ref{fig:F},
we see that the fidelity remains approximately unity even for times
near the middle of the protocol owing to the still large Zeeman gap.
This implies one can start from an initial product state and yet attain
large fidelities for the Kitaev ground state at the end of protocol. 

\begin{figure}
\includegraphics[width=1\columnwidth]{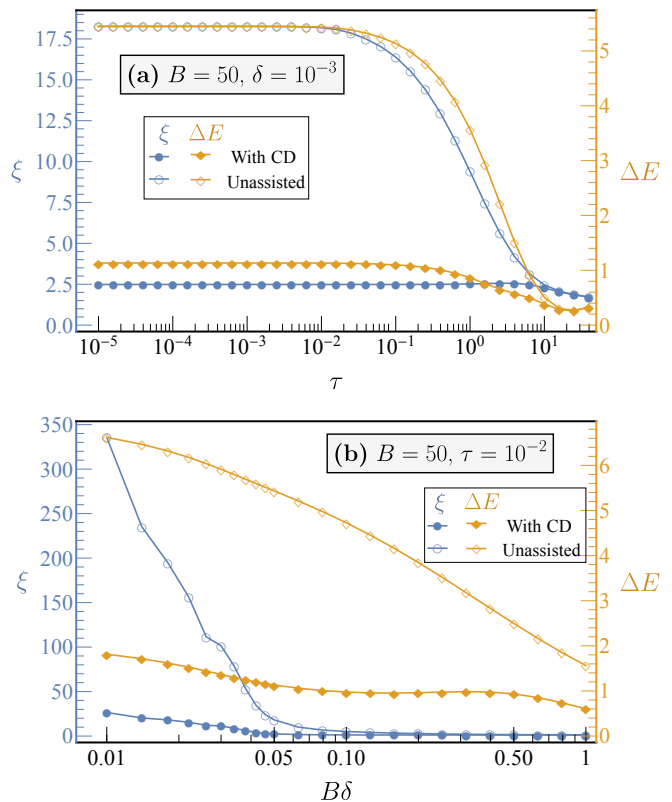}\caption{\label{fig:xi}Support size, $\xi$ and change in the energy of the
evolved state from the ground state energy, $\Delta E=\left\langle E\right\rangle -E_{GS}$
are plotted as a function of (a) protocol duration, $\tau$ and (b)
final Zeeman field, $B\delta$. In (a) CD assisted support size, $\xi_{\text{CD}}\sim1$
while for naive protocol, $\xi\sim10$ for a wide stretch of $\tau$values.
The CD assisted quantum state obtained is localised in the bottom
part of the energy spectrum is confirmed by the energy difference,
$\Delta E$. Again as in Fig. \ref{fig:FgammaW}a , the naive driving
of Hamiltonian parameters start coinciding with the transitionless
driving at $\tau\sim10.$ Even for very small Zeeman fields, (b) shows
the support size, $\xi_{\text{CD}}$ is still localised to $\sim10$
states only in a many body Hilbert space of dimension $2^{24}$ while
$\xi$ for the unassisted case shows a very rapid variation.}
\end{figure}

We next compare the usefulness of the quantum state obtained via the
two protocols by studying their entanglement entropy and plaquette
fluxes. The Kitaev ground state is associated with a finite topological
entanglement entropy, which is the part of the von Neumann bipartite
entropy, $S_{A}=\text{Tr}\rho_{A}\log\rho_{A}=\alpha L-\gamma$, remaining
after subtracting the area law contribution. Here $L$ is the perimeter
of a 2D subsystem $A$ whose bipartite entanglement entropy is $S_{A}.$
For the Kitaev ground state, $\gamma=\ln2.$ Appropriately choosing
four partitions of the lattice (see Fig. \ref{fig:lattice}) and taking
a linear combination of entropies of three of the partitions yields
$\gamma$, which is free from the boundary term \citep{Kitaev_Preskill}:
\begin{equation}
-\gamma=S_{A}+S_{B}+S_{C}-S_{AB}-S_{BC}-S_{AC}+S_{ABC}.
\end{equation}
 Finite topological entanglement entropy ensures the quantum state
is fault-tolerant to local disturbances. In addition to the entanglement
entropy, the plaquette flux operator $\hat{W}$ is another characteristic
quantity in pure Kitaev system. With reference to Fig. \ref{fig:lattice},
we choose the plaquette, $p=\{1,4,9,13,10,5\}$ for computing expectation
value of the flux operator defined as $\hat{W}_{p}=\sigma_{1}^{z}\sigma_{4}^{x}\sigma_{9}^{y}\sigma_{13}^{z}\sigma_{10}^{x}\sigma_{5}^{y}$.
The flux operator commutes with the Kitaev Hamiltonian, $\hat{H}_{\text{k}}$
and has eigenvalues $W=\pm1.$ Ground state is characterised by $W=1.$
In Fig. \ref{fig:FgammaW} we show the dependence of these quantities
on (a) $\tau$ and (b) final Zeeman field, $B\delta.$ The green coloured
line in Fig. \ref{fig:FgammaW} shows the value of these quantities
corresponding to the ground state of the final Hamiltonian for comparison.
We find that for $\tau\lesssim1,$ the topological entanglement entropy
of the quantum state evolved using transitionless driving is an order
of magnitude higher than that of the state evolved without it. A finite
nonzero $\gamma$ implies the state is useful for quantum computation.
Similarly, CD assisted protocol yields values of $W$ closer to unity
compared to unassisted driving. Because of better fidelity through
counterdiabatic approach, the final time-evolved quantum state stays
close to the true ground state. Quantitatively, this can be expressed
in terms of the support size $\xi$ of the quantum state (obtained
at the end of the protocol) in the many-body Hilbert space of the
exact eigenstates of the final Hamiltonian. The support size $\xi$
is defined as $\xi^{-1}=\sum_{i}\left|a_{i}\right|^{4}$ , where $a_{i}$
is the overlap of the quantum state with the many-body eigenstates
$|i\rangle.$ We observe in Fig. \ref{fig:xi}a support sizes of $2.5$
states or less when CD assisted for a wide range of protocol durations
while unassisted driving gives support sizes of the order of 10 states
or more. In CD protocol, $\xi$ shows a relatively slower variation
on changing the final Zeeman field as opposed to unassisted protocol
where $\xi$ rapidly rises for decreasing $B\delta$ as shown in Fig.
\ref{fig:xi}b. Along with $\xi$ in Fig. \ref{fig:xi} we have plotted
the energy difference of the obtained quantum state with the pure
ground state. We find the energy difference is small for the CD case.
We thus claim that the CD aided approach yields the quantum state
with high localisation (the dimensionality of the many-body Hilbert
space for our 24-spin cluster is $2^{24}$) near the ground state,
with the ground state having the maximum share equal to its fidelity.

\section{Discussion \label{sec:Discussion}}

The Kitaev ground state is useful for universal quantum computing
and storage due to its long-range topological order and anyonic excitations.
We demonstrated using a counterdiabatic strategy, the feasibility
of preparing the Kitaev ground state (in the presence of a small Zeeman
field sufficient for introducing a bulk spectral gap) with high fidelity
on time scales significantly smaller than that permitted by the adiabatic
theorem. Our proposed method relies on local two-body interactions
which makes it practicable for implementation. Features such as topological
entanglement entropy and nonzero flux expectation were shown to be
preserved much better than unassisted protocols in the same time duration.
Remarkably, these properties are found to be rather insensitive to
the (short) duration of the protocol and instead depend much more
on the parameters of the final Hamiltonian. To further increase the
fidelity and other properties, one must include higher order (and
more nonlocal) terms neglected in our approximate CD protocol.
\begin{acknowledgments}
We thank Aman Kumar for sharing the DMRG result of energy gaps. 
\end{acknowledgments}

\appendix
\begin{widetext}

\section{Gauge Potential, $\hat{A}_{\lambda}$ \label{sec:AppendixA}}

We sketch the steps involved to obtain the counterdiabatic perturbation
to the Hamiltonian $\hat{H}_{0}$. For 2-body local interaction, we
expand eq. (\ref{eq:A}) to leading order:

\begin{equation}
\hat{A}_{\lambda}^{(1)}=i\alpha_{1}[\hat{H}_{0},\partial_{\lambda}\hat{H}_{0}]=i\alpha_{1}[\hat{H}_{\text{k}},\hat{H}_{\text{m}}].\label{eq:A-1}
\end{equation}
 Let the Zeeman field coupling constant $\boldsymbol{B}=(B_{x},B_{y},B_{z}).$
In our analysis, we considered the field along $[111]-$ direction
with $B_{x}=B_{y}=B_{z}=B$. The commutator in eq. (\ref{eq:A1})
can be written as a sum of 3 terms $[\hat{H}_{\text{k}},\hat{H}_{\text{m}}]=\hat{\mathcal{X}}+\hat{\mathcal{Y}}+\hat{\mathcal{Z}}$
which are given by
\begin{equation}
\hat{\mathcal{X}}=2iB_{x}J\left\{ \sum_{\left\langle jk\right\rangle _{z-\text{{link}}}}\left(\sigma_{j}^{y}\sigma_{k}^{z}+\sigma_{j}^{z}\sigma_{k}^{y}\right)-\sum_{\left\langle jk\right\rangle _{y-\text{{link}}}}\left(\sigma_{j}^{y}\sigma_{k}^{z}+\sigma_{j}^{z}\sigma_{k}^{y}\right)\right\} ,
\end{equation}

\begin{equation}
\hat{\mathcal{Y}}=2iB_{y}J\left\{ \sum_{\left\langle jk\right\rangle _{x-\text{{link}}}}\left(\sigma_{j}^{z}\sigma_{k}^{x}+\sigma_{j}^{x}\sigma_{k}^{z}\right)-\sum_{\left\langle jk\right\rangle _{z-\text{{link}}}}\left(\sigma_{j}^{z}\sigma_{k}^{x}+\sigma_{j}^{x}\sigma_{k}^{z}\right)\right\} ,
\end{equation}

\begin{equation}
\hat{\mathcal{Z}}=2iB_{z}J\left\{ \sum_{\left\langle jk\right\rangle _{y-\text{{link}}}}\left(\sigma_{j}^{x}\sigma_{k}^{y}+\sigma_{j}^{y}\sigma_{k}^{x}\right)-\sum_{\left\langle jk\right\rangle _{x-\text{{link}}}}\left(\sigma_{j}^{x}\sigma_{k}^{y}+\sigma_{j}^{y}\sigma_{k}^{x}\right)\right\} .
\end{equation}
The variational parameter $\alpha_{1}$ is found by constructing the
operator $\hat{G}$ as defined in eq. (\ref{eq:G}) and minimizing
$S=\text{Tr }\hat{G^{2}}$. The leading order of $\hat{G}$ can be
expressed as
\begin{align}
\hat{G}^{(1)} & =\partial_{\lambda}\hat{H}_{0}-i\left[\hat{H}_{0},\hat{A}_{\lambda}^{(1)}\right]\nonumber \\
 & =\hat{H}_{\text{m}}-\alpha_{1}\left(\lambda+\delta\right)\left[\hat{H}_{\text{m}},\left[\hat{H}_{\text{m}},\hat{H}_{\text{k}}\right]\right]-\alpha_{1}\left[\hat{H}_{\text{k}},\left[\hat{H}_{\text{m}},\hat{H}_{\text{k}}\right]\right].\nonumber \\
\end{align}
The minimization condition $\delta S/\delta\alpha_{1}=0$ yields
\begin{equation}
\alpha_{1}=\frac{\left(\lambda+\delta\right)\text{Tr}\left(\hat{H}_{\text{m}}\left[\hat{H}_{\text{m}},\left[\hat{H}_{\text{m}},\hat{H}_{\text{k}}\right]\right]\right)+\text{Tr}\left(\hat{H}_{\text{m}}\left[\hat{H}_{\text{k}},\left[\hat{H}_{\text{m}},\hat{H}_{\text{k}}\right]\right]\right)}{\left(\lambda+\delta\right)^{2}\text{Tr}\left(\left[\hat{H}_{\text{m}},\left[\hat{H}_{\text{m}},\hat{H}_{\text{k}}\right]\right]^{2}\right)+\text{Tr}\left(\left[\hat{H}_{\text{k}},\left[\hat{H}_{\text{m}},\hat{H}_{\text{k}}\right]\right]^{2}\right)+2\left(\lambda+\delta\right)\text{Tr}\left(\left[\hat{H}_{\text{m}},\left[\hat{H}_{\text{m}},\hat{H}_{\text{k}}\right]\right]\left[\hat{H}_{\text{k}},\left[\hat{H}_{\text{m}},\hat{H}_{\text{k}}\right]\right]\right)}
\end{equation}
The traces are evaluated numerically resulting in 
\begin{equation}
\alpha_{1}=\frac{-1/4}{9\left(\lambda+\delta\right)^{2}B^{2}+5J^{2}}
\end{equation}
Substituting $\alpha_{1}$ in eq. (\ref{eq:A-1}) gives the gauge
potential $\hat{A}_{\lambda}^{(1)}.$

\section{Energy gap \label{sec:AppendixB}}

Higher fidelities in CD asssisted protocols are aided by the mass
gap present in the Kitaev system in presence of a Zeeman field. Here,
we illustrate this gap is not a finite size effect. Figure (\ref{fig:gap})
shows the comparison of energy gap between ground state and first
excited state for the Hamiltonian, $\hat{H}=\hat{H}_{\text{m}}+\hat{H}_{\text{k}}$
in $24-$sites and $144-$sites lattice. All $24-$sites calculations
are performed via exact diagonalistion in Quspin while the larger
$144-$sites energy gap calculation is done using finite size DMRG.
We note the gap coincides for the 2 cases for a wide range of magnetic
field encountered in the protocol (see inset of Fig. (\ref{fig:F}b)).
\begin{figure}
\includegraphics[width=0.5\columnwidth]{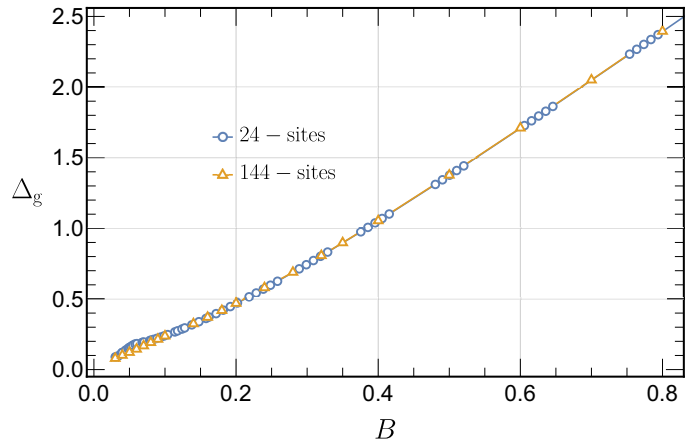}

\caption{\label{fig:gap}Ground state energy gap for the Hamiltonian, $\hat{H}=\hat{H}_{\text{m}}+\hat{H}_{\text{k}}$
as calculated for $24-$sites using exact diagonalisation and $144-$sites
via DMRG as a function of Zeeman field coupling $B$ is shown. The
2 cases coincide for a wide range of magnetic field values encountered
in the naive as well as CD assisted protocol (see inset of Fig. (\ref{fig:F}b)).}
\end{figure}

\end{widetext}

\bibliographystyle{apsrev4-2}
\bibliography{cd}

\begin{thebibliography}{31}%
\makeatletter
\providecommand \@ifxundefined [1]{%
 \@ifx{#1\undefined}
}%
\providecommand \@ifnum [1]{%
 \ifnum #1\expandafter \@firstoftwo
 \else \expandafter \@secondoftwo
 \fi
}%
\providecommand \@ifx [1]{%
 \ifx #1\expandafter \@firstoftwo
 \else \expandafter \@secondoftwo
 \fi
}%
\providecommand \natexlab [1]{#1}%
\providecommand \enquote  [1]{``#1''}%
\providecommand \bibnamefont  [1]{#1}%
\providecommand \bibfnamefont [1]{#1}%
\providecommand \citenamefont [1]{#1}%
\providecommand \href@noop [0]{\@secondoftwo}%
\providecommand \href [0]{\begingroup \@sanitize@url \@href}%
\providecommand \@href[1]{\@@startlink{#1}\@@href}%
\providecommand \@@href[1]{\endgroup#1\@@endlink}%
\providecommand \@sanitize@url [0]{\catcode `\\12\catcode `\$12\catcode
  `\&12\catcode `\#12\catcode `\^12\catcode `\_12\catcode `\%12\relax}%
\providecommand \@@startlink[1]{}%
\providecommand \@@endlink[0]{}%
\providecommand \url  [0]{\begingroup\@sanitize@url \@url }%
\providecommand \@url [1]{\endgroup\@href {#1}{\urlprefix }}%
\providecommand \urlprefix  [0]{URL }%
\providecommand \Eprint [0]{\href }%
\providecommand \doibase [0]{https://doi.org/}%
\providecommand \selectlanguage [0]{\@gobble}%
\providecommand \bibinfo  [0]{\@secondoftwo}%
\providecommand \bibfield  [0]{\@secondoftwo}%
\providecommand \translation [1]{[#1]}%
\providecommand \BibitemOpen [0]{}%
\providecommand \bibitemStop [0]{}%
\providecommand \bibitemNoStop [0]{.\EOS\space}%
\providecommand \EOS [0]{\spacefactor3000\relax}%
\providecommand \BibitemShut  [1]{\csname bibitem#1\endcsname}%
\let\auto@bib@innerbib\@empty
\bibitem [{\citenamefont {Kitaev}(2003)}]{kitaev2003fault}%
  \BibitemOpen
  \bibfield  {author} {\bibinfo {author} {\bibfnamefont {A.~Y.}\ \bibnamefont
  {Kitaev}},\ }\href@noop {} {\bibfield  {journal} {\bibinfo  {journal} {Annals
  of Physics}\ }\textbf {\bibinfo {volume} {303}},\ \bibinfo {pages} {2}
  (\bibinfo {year} {2003})}\BibitemShut {NoStop}%
\bibitem [{\citenamefont {Gross}\ \emph {et~al.}(2009)\citenamefont {Gross},
  \citenamefont {Flammia},\ and\ \citenamefont {Eisert}}]{Gross_too_entangled}%
  \BibitemOpen
  \bibfield  {author} {\bibinfo {author} {\bibfnamefont {D.}~\bibnamefont
  {Gross}}, \bibinfo {author} {\bibfnamefont {S.~T.}\ \bibnamefont {Flammia}},\
  and\ \bibinfo {author} {\bibfnamefont {J.}~\bibnamefont {Eisert}},\ }\href
  {https://doi.org/10.1103/PhysRevLett.102.190501} {\bibfield  {journal}
  {\bibinfo  {journal} {Phys. Rev. Lett.}\ }\textbf {\bibinfo {volume} {102}},\
  \bibinfo {pages} {190501} (\bibinfo {year} {2009})}\BibitemShut {NoStop}%
\bibitem [{\citenamefont {He}\ \emph {et~al.}(2016)\citenamefont {He},
  \citenamefont {Su}, \citenamefont {Wang}, \citenamefont {Sun}, \citenamefont
  {Bai}, \citenamefont {Zhu}, \citenamefont {Wang},\ and\ \citenamefont
  {Zhang}}]{TwoatomEntaglement_prep_nature}%
  \BibitemOpen
  \bibfield  {author} {\bibinfo {author} {\bibfnamefont {S.}~\bibnamefont
  {He}}, \bibinfo {author} {\bibfnamefont {S.-L.}\ \bibnamefont {Su}}, \bibinfo
  {author} {\bibfnamefont {D.-Y.}\ \bibnamefont {Wang}}, \bibinfo {author}
  {\bibfnamefont {W.-M.}\ \bibnamefont {Sun}}, \bibinfo {author} {\bibfnamefont
  {C.-H.}\ \bibnamefont {Bai}}, \bibinfo {author} {\bibfnamefont {A.-D.}\
  \bibnamefont {Zhu}}, \bibinfo {author} {\bibfnamefont {H.-F.}\ \bibnamefont
  {Wang}},\ and\ \bibinfo {author} {\bibfnamefont {S.}~\bibnamefont {Zhang}},\
  }\href@noop {} {\bibfield  {journal} {\bibinfo  {journal} {Scientific
  reports}\ }\textbf {\bibinfo {volume} {6}},\ \bibinfo {pages} {1} (\bibinfo
  {year} {2016})}\BibitemShut {NoStop}%
\bibitem [{\citenamefont {Bandyopadhyay}\ and\ \citenamefont
  {Dutta}(2020)}]{ChernInsulator_prep_prb}%
  \BibitemOpen
  \bibfield  {author} {\bibinfo {author} {\bibfnamefont {S.}~\bibnamefont
  {Bandyopadhyay}}\ and\ \bibinfo {author} {\bibfnamefont {A.}~\bibnamefont
  {Dutta}},\ }\href {https://doi.org/10.1103/PhysRevB.102.094301} {\bibfield
  {journal} {\bibinfo  {journal} {Phys. Rev. B}\ }\textbf {\bibinfo {volume}
  {102}},\ \bibinfo {pages} {094301} (\bibinfo {year} {2020})}\BibitemShut
  {NoStop}%
\bibitem [{\citenamefont {Hamma}\ and\ \citenamefont
  {Lidar}(2008)}]{adiab_topo_prep_prl}%
  \BibitemOpen
  \bibfield  {author} {\bibinfo {author} {\bibfnamefont {A.}~\bibnamefont
  {Hamma}}\ and\ \bibinfo {author} {\bibfnamefont {D.~A.}\ \bibnamefont
  {Lidar}},\ }\href {https://doi.org/10.1103/PhysRevLett.100.030502} {\bibfield
   {journal} {\bibinfo  {journal} {Phys. Rev. Lett.}\ }\textbf {\bibinfo
  {volume} {100}},\ \bibinfo {pages} {030502} (\bibinfo {year}
  {2008})}\BibitemShut {NoStop}%
\bibitem [{\citenamefont {Santos}\ and\ \citenamefont
  {Sarandy}(2017)}]{decoherence_cd_iop}%
  \BibitemOpen
  \bibfield  {author} {\bibinfo {author} {\bibfnamefont {A.~C.}\ \bibnamefont
  {Santos}}\ and\ \bibinfo {author} {\bibfnamefont {M.~S.}\ \bibnamefont
  {Sarandy}},\ }\href {https://doi.org/10.1088/1751-8121/aa96f1} {\bibfield
  {journal} {\bibinfo  {journal} {Journal of Physics A: Mathematical and
  Theoretical}\ }\textbf {\bibinfo {volume} {51}},\ \bibinfo {pages} {025301}
  (\bibinfo {year} {2017})}\BibitemShut {NoStop}%
\bibitem [{\citenamefont {Gu\'ery-Odelin}\ \emph {et~al.}(2019)\citenamefont
  {Gu\'ery-Odelin}, \citenamefont {Ruschhaupt}, \citenamefont {Kiely},
  \citenamefont {Torrontegui}, \citenamefont {Mart\'{\i}nez-Garaot},\ and\
  \citenamefont {Muga}}]{STA_concepts}%
  \BibitemOpen
  \bibfield  {author} {\bibinfo {author} {\bibfnamefont {D.}~\bibnamefont
  {Gu\'ery-Odelin}}, \bibinfo {author} {\bibfnamefont {A.}~\bibnamefont
  {Ruschhaupt}}, \bibinfo {author} {\bibfnamefont {A.}~\bibnamefont {Kiely}},
  \bibinfo {author} {\bibfnamefont {E.}~\bibnamefont {Torrontegui}}, \bibinfo
  {author} {\bibfnamefont {S.}~\bibnamefont {Mart\'{\i}nez-Garaot}},\ and\
  \bibinfo {author} {\bibfnamefont {J.~G.}\ \bibnamefont {Muga}},\ }\href
  {https://doi.org/10.1103/RevModPhys.91.045001} {\bibfield  {journal}
  {\bibinfo  {journal} {Rev. Mod. Phys.}\ }\textbf {\bibinfo {volume} {91}},\
  \bibinfo {pages} {045001} (\bibinfo {year} {2019})}\BibitemShut {NoStop}%
\bibitem [{\citenamefont {Kolodrubetz}\ \emph {et~al.}(2017)\citenamefont
  {Kolodrubetz}, \citenamefont {Sels}, \citenamefont {Mehta},\ and\
  \citenamefont {Polkovnikov}}]{Polkovnikov_lectures}%
  \BibitemOpen
  \bibfield  {author} {\bibinfo {author} {\bibfnamefont {M.}~\bibnamefont
  {Kolodrubetz}}, \bibinfo {author} {\bibfnamefont {D.}~\bibnamefont {Sels}},
  \bibinfo {author} {\bibfnamefont {P.}~\bibnamefont {Mehta}},\ and\ \bibinfo
  {author} {\bibfnamefont {A.}~\bibnamefont {Polkovnikov}},\ }\href
  {https://doi.org/https://doi.org/10.1016/j.physrep.2017.07.001} {\bibfield
  {journal} {\bibinfo  {journal} {Physics Reports}\ }\textbf {\bibinfo {volume}
  {697}},\ \bibinfo {pages} {1} (\bibinfo {year} {2017})}\BibitemShut {NoStop}%
\bibitem [{\citenamefont {Chen}\ \emph {et~al.}(2010)\citenamefont {Chen},
  \citenamefont {Ruschhaupt}, \citenamefont {Schmidt}, \citenamefont {del
  Campo}, \citenamefont {Gu\'ery-Odelin},\ and\ \citenamefont
  {Muga}}]{SHM_STA_prl}%
  \BibitemOpen
  \bibfield  {author} {\bibinfo {author} {\bibfnamefont {X.}~\bibnamefont
  {Chen}}, \bibinfo {author} {\bibfnamefont {A.}~\bibnamefont {Ruschhaupt}},
  \bibinfo {author} {\bibfnamefont {S.}~\bibnamefont {Schmidt}}, \bibinfo
  {author} {\bibfnamefont {A.}~\bibnamefont {del Campo}}, \bibinfo {author}
  {\bibfnamefont {D.}~\bibnamefont {Gu\'ery-Odelin}},\ and\ \bibinfo {author}
  {\bibfnamefont {J.~G.}\ \bibnamefont {Muga}},\ }\href
  {https://doi.org/10.1103/PhysRevLett.104.063002} {\bibfield  {journal}
  {\bibinfo  {journal} {Phys. Rev. Lett.}\ }\textbf {\bibinfo {volume} {104}},\
  \bibinfo {pages} {063002} (\bibinfo {year} {2010})}\BibitemShut {NoStop}%
\bibitem [{\citenamefont {Deffner}\ \emph {et~al.}(2014)\citenamefont
  {Deffner}, \citenamefont {Jarzynski},\ and\ \citenamefont {del
  Campo}}]{Jarzynski_prx}%
  \BibitemOpen
  \bibfield  {author} {\bibinfo {author} {\bibfnamefont {S.}~\bibnamefont
  {Deffner}}, \bibinfo {author} {\bibfnamefont {C.}~\bibnamefont {Jarzynski}},\
  and\ \bibinfo {author} {\bibfnamefont {A.}~\bibnamefont {del Campo}},\ }\href
  {https://doi.org/10.1103/PhysRevX.4.021013} {\bibfield  {journal} {\bibinfo
  {journal} {Phys. Rev. X}\ }\textbf {\bibinfo {volume} {4}},\ \bibinfo {pages}
  {021013} (\bibinfo {year} {2014})}\BibitemShut {NoStop}%
\bibitem [{\citenamefont {Jarzynski}(2013)}]{Jarzynski_pra}%
  \BibitemOpen
  \bibfield  {author} {\bibinfo {author} {\bibfnamefont {C.}~\bibnamefont
  {Jarzynski}},\ }\href {https://doi.org/10.1103/PhysRevA.88.040101} {\bibfield
   {journal} {\bibinfo  {journal} {Phys. Rev. A}\ }\textbf {\bibinfo {volume}
  {88}},\ \bibinfo {pages} {040101} (\bibinfo {year} {2013})}\BibitemShut
  {NoStop}%
\bibitem [{\citenamefont {Berry}(2009)}]{Berry_iop}%
  \BibitemOpen
  \bibfield  {author} {\bibinfo {author} {\bibfnamefont {M.~V.}\ \bibnamefont
  {Berry}},\ }\href@noop {} {\bibfield  {journal} {\bibinfo  {journal} {Journal
  of Physics A: Mathematical and Theoretical}\ }\textbf {\bibinfo {volume}
  {42}},\ \bibinfo {pages} {365303} (\bibinfo {year} {2009})}\BibitemShut
  {NoStop}%
\bibitem [{\citenamefont {Duncan}\ and\ \citenamefont
  {Del~Campo}(2018)}]{CDBO_STA_iop}%
  \BibitemOpen
  \bibfield  {author} {\bibinfo {author} {\bibfnamefont {C.~W.}\ \bibnamefont
  {Duncan}}\ and\ \bibinfo {author} {\bibfnamefont {A.}~\bibnamefont
  {Del~Campo}},\ }\href@noop {} {\bibfield  {journal} {\bibinfo  {journal} {New
  Journal of Physics}\ }\textbf {\bibinfo {volume} {20}},\ \bibinfo {pages}
  {085003} (\bibinfo {year} {2018})}\BibitemShut {NoStop}%
\bibitem [{\citenamefont {Hartmann}\ and\ \citenamefont
  {Lechner}(2019)}]{Hartmann_iop}%
  \BibitemOpen
  \bibfield  {author} {\bibinfo {author} {\bibfnamefont {A.}~\bibnamefont
  {Hartmann}}\ and\ \bibinfo {author} {\bibfnamefont {W.}~\bibnamefont
  {Lechner}},\ }\href {https://doi.org/10.1088/1367-2630/ab14a0} {\bibfield
  {journal} {\bibinfo  {journal} {New Journal of Physics}\ }\textbf {\bibinfo
  {volume} {21}},\ \bibinfo {pages} {043025} (\bibinfo {year}
  {2019})}\BibitemShut {NoStop}%
\bibitem [{\citenamefont {Sels}\ and\ \citenamefont
  {Polkovnikov}(2017)}]{Polkovnikov_waiter_pnas}%
  \BibitemOpen
  \bibfield  {author} {\bibinfo {author} {\bibfnamefont {D.}~\bibnamefont
  {Sels}}\ and\ \bibinfo {author} {\bibfnamefont {A.}~\bibnamefont
  {Polkovnikov}},\ }\href@noop {} {\bibfield  {journal} {\bibinfo  {journal}
  {Proceedings of the National Academy of Sciences}\ }\textbf {\bibinfo
  {volume} {114}},\ \bibinfo {pages} {E3909} (\bibinfo {year}
  {2017})}\BibitemShut {NoStop}%
\bibitem [{\citenamefont {Meier}\ \emph {et~al.}(2020)\citenamefont {Meier},
  \citenamefont {Ngan}, \citenamefont {Sels},\ and\ \citenamefont
  {Gadway}}]{ninesite_cd_prr}%
  \BibitemOpen
  \bibfield  {author} {\bibinfo {author} {\bibfnamefont {E.~J.}\ \bibnamefont
  {Meier}}, \bibinfo {author} {\bibfnamefont {K.}~\bibnamefont {Ngan}},
  \bibinfo {author} {\bibfnamefont {D.}~\bibnamefont {Sels}},\ and\ \bibinfo
  {author} {\bibfnamefont {B.}~\bibnamefont {Gadway}},\ }\href
  {https://doi.org/10.1103/PhysRevResearch.2.043201} {\bibfield  {journal}
  {\bibinfo  {journal} {Phys. Rev. Research}\ }\textbf {\bibinfo {volume}
  {2}},\ \bibinfo {pages} {043201} (\bibinfo {year} {2020})}\BibitemShut
  {NoStop}%
\bibitem [{\citenamefont {Kitaev}(2006)}]{kitaev2006anyons}%
  \BibitemOpen
  \bibfield  {author} {\bibinfo {author} {\bibfnamefont {A.}~\bibnamefont
  {Kitaev}},\ }\href@noop {} {\bibfield  {journal} {\bibinfo  {journal} {Annals
  of Physics}\ }\textbf {\bibinfo {volume} {321}},\ \bibinfo {pages} {2}
  (\bibinfo {year} {2006})}\BibitemShut {NoStop}%
\bibitem [{\citenamefont {Kitaev}\ and\ \citenamefont
  {Preskill}(2006)}]{Kitaev_Preskill}%
  \BibitemOpen
  \bibfield  {author} {\bibinfo {author} {\bibfnamefont {A.}~\bibnamefont
  {Kitaev}}\ and\ \bibinfo {author} {\bibfnamefont {J.}~\bibnamefont
  {Preskill}},\ }\href {https://doi.org/10.1103/PhysRevLett.96.110404}
  {\bibfield  {journal} {\bibinfo  {journal} {Phys. Rev. Lett.}\ }\textbf
  {\bibinfo {volume} {96}},\ \bibinfo {pages} {110404} (\bibinfo {year}
  {2006})}\BibitemShut {NoStop}%
\bibitem [{\citenamefont {Lloyd}(2002)}]{lloyd2002quantum}%
  \BibitemOpen
  \bibfield  {author} {\bibinfo {author} {\bibfnamefont {S.}~\bibnamefont
  {Lloyd}},\ }\href@noop {} {\bibfield  {journal} {\bibinfo  {journal} {Quantum
  Information Processing}\ }\textbf {\bibinfo {volume} {1}},\ \bibinfo {pages}
  {13} (\bibinfo {year} {2002})}\BibitemShut {NoStop}%
\bibitem [{\citenamefont {Altshuler}\ \emph {et~al.}(2010)\citenamefont
  {Altshuler}, \citenamefont {Krovi},\ and\ \citenamefont
  {Roland}}]{altshuler_anderson_pnas}%
  \BibitemOpen
  \bibfield  {author} {\bibinfo {author} {\bibfnamefont {B.}~\bibnamefont
  {Altshuler}}, \bibinfo {author} {\bibfnamefont {H.}~\bibnamefont {Krovi}},\
  and\ \bibinfo {author} {\bibfnamefont {J.}~\bibnamefont {Roland}},\
  }\href@noop {} {\bibfield  {journal} {\bibinfo  {journal} {Proceedings of the
  National Academy of Sciences}\ }\textbf {\bibinfo {volume} {107}},\ \bibinfo
  {pages} {12446} (\bibinfo {year} {2010})}\BibitemShut {NoStop}%
\bibitem [{\citenamefont {Gohlke}\ \emph {et~al.}(2018)\citenamefont {Gohlke},
  \citenamefont {Moessner},\ and\ \citenamefont
  {Pollmann}}]{magnetic_Kitaev_prb}%
  \BibitemOpen
  \bibfield  {author} {\bibinfo {author} {\bibfnamefont {M.}~\bibnamefont
  {Gohlke}}, \bibinfo {author} {\bibfnamefont {R.}~\bibnamefont {Moessner}},\
  and\ \bibinfo {author} {\bibfnamefont {F.}~\bibnamefont {Pollmann}},\ }\href
  {https://doi.org/10.1103/PhysRevB.98.014418} {\bibfield  {journal} {\bibinfo
  {journal} {Phys. Rev. B}\ }\textbf {\bibinfo {volume} {98}},\ \bibinfo
  {pages} {014418} (\bibinfo {year} {2018})}\BibitemShut {NoStop}%
\bibitem [{\citenamefont {Rau}\ and\ \citenamefont {Kee}(2014)}]{trigonal1}%
  \BibitemOpen
  \bibfield  {author} {\bibinfo {author} {\bibfnamefont {J.~G.}\ \bibnamefont
  {Rau}}\ and\ \bibinfo {author} {\bibfnamefont {H.-Y.}\ \bibnamefont {Kee}},\
  }\href@noop {} {\bibfield  {journal} {\bibinfo  {journal} {arXiv:1408.4811}\
  } (\bibinfo {year} {2014})}\BibitemShut {NoStop}%
\bibitem [{\citenamefont {Takikawa}\ and\ \citenamefont
  {Fujimoto}(2020)}]{trigonal2}%
  \BibitemOpen
  \bibfield  {author} {\bibinfo {author} {\bibfnamefont {D.}~\bibnamefont
  {Takikawa}}\ and\ \bibinfo {author} {\bibfnamefont {S.}~\bibnamefont
  {Fujimoto}},\ }\href {https://doi.org/10.1103/PhysRevB.102.174414} {\bibfield
   {journal} {\bibinfo  {journal} {Phys. Rev. B}\ }\textbf {\bibinfo {volume}
  {102}},\ \bibinfo {pages} {174414} (\bibinfo {year} {2020})}\BibitemShut
  {NoStop}%
\bibitem [{\citenamefont {Takikawa}\ and\ \citenamefont
  {Fujimoto}(2019)}]{trigonal3}%
  \BibitemOpen
  \bibfield  {author} {\bibinfo {author} {\bibfnamefont {D.}~\bibnamefont
  {Takikawa}}\ and\ \bibinfo {author} {\bibfnamefont {S.}~\bibnamefont
  {Fujimoto}},\ }\href {https://doi.org/10.1103/PhysRevB.99.224409} {\bibfield
  {journal} {\bibinfo  {journal} {Phys. Rev. B}\ }\textbf {\bibinfo {volume}
  {99}},\ \bibinfo {pages} {224409} (\bibinfo {year} {2019})}\BibitemShut
  {NoStop}%
\bibitem [{\citenamefont {Sameti}\ and\ \citenamefont
  {Hartmann}(2019)}]{quantum_simulation_PhysRevA.99.012333}%
  \BibitemOpen
  \bibfield  {author} {\bibinfo {author} {\bibfnamefont {M.}~\bibnamefont
  {Sameti}}\ and\ \bibinfo {author} {\bibfnamefont {M.~J.}\ \bibnamefont
  {Hartmann}},\ }\href {https://doi.org/10.1103/PhysRevA.99.012333} {\bibfield
  {journal} {\bibinfo  {journal} {Phys. Rev. A}\ }\textbf {\bibinfo {volume}
  {99}},\ \bibinfo {pages} {012333} (\bibinfo {year} {2019})}\BibitemShut
  {NoStop}%
\bibitem [{\citenamefont {You}\ \emph {et~al.}(2010)\citenamefont {You},
  \citenamefont {Shi}, \citenamefont {Hu},\ and\ \citenamefont
  {Nori}}]{quantum_simulation_PhysRevB.81.014505}%
  \BibitemOpen
  \bibfield  {author} {\bibinfo {author} {\bibfnamefont {J.~Q.}\ \bibnamefont
  {You}}, \bibinfo {author} {\bibfnamefont {X.-F.}\ \bibnamefont {Shi}},
  \bibinfo {author} {\bibfnamefont {X.}~\bibnamefont {Hu}},\ and\ \bibinfo
  {author} {\bibfnamefont {F.}~\bibnamefont {Nori}},\ }\href
  {https://doi.org/10.1103/PhysRevB.81.014505} {\bibfield  {journal} {\bibinfo
  {journal} {Phys. Rev. B}\ }\textbf {\bibinfo {volume} {81}},\ \bibinfo
  {pages} {014505} (\bibinfo {year} {2010})}\BibitemShut {NoStop}%
\bibitem [{\citenamefont {Claeys}\ \emph {et~al.}(2019)\citenamefont {Claeys},
  \citenamefont {Pandey}, \citenamefont {Sels},\ and\ \citenamefont
  {Polkovnikov}}]{Polkovnikov_floquet_cd_prl}%
  \BibitemOpen
  \bibfield  {author} {\bibinfo {author} {\bibfnamefont {P.~W.}\ \bibnamefont
  {Claeys}}, \bibinfo {author} {\bibfnamefont {M.}~\bibnamefont {Pandey}},
  \bibinfo {author} {\bibfnamefont {D.}~\bibnamefont {Sels}},\ and\ \bibinfo
  {author} {\bibfnamefont {A.}~\bibnamefont {Polkovnikov}},\ }\href
  {https://doi.org/10.1103/PhysRevLett.123.090602} {\bibfield  {journal}
  {\bibinfo  {journal} {Phys. Rev. Lett.}\ }\textbf {\bibinfo {volume} {123}},\
  \bibinfo {pages} {090602} (\bibinfo {year} {2019})}\BibitemShut {NoStop}%
\bibitem [{\citenamefont {Kyaw}\ and\ \citenamefont {Kwek}(2018)}]{kyaw_iop}%
  \BibitemOpen
  \bibfield  {author} {\bibinfo {author} {\bibfnamefont {T.~H.}\ \bibnamefont
  {Kyaw}}\ and\ \bibinfo {author} {\bibfnamefont {L.-C.}\ \bibnamefont
  {Kwek}},\ }\href@noop {} {\bibfield  {journal} {\bibinfo  {journal} {New
  Journal of Physics}\ }\textbf {\bibinfo {volume} {20}},\ \bibinfo {pages}
  {045007} (\bibinfo {year} {2018})}\BibitemShut {NoStop}%
\bibitem [{\citenamefont {Weinberg}\ and\ \citenamefont
  {Bukov}(2017)}]{quspin2017}%
  \BibitemOpen
  \bibfield  {author} {\bibinfo {author} {\bibfnamefont {P.}~\bibnamefont
  {Weinberg}}\ and\ \bibinfo {author} {\bibfnamefont {M.}~\bibnamefont
  {Bukov}},\ }\href {https://doi.org/10.21468/SciPostPhys.2.1.003} {\bibfield
  {journal} {\bibinfo  {journal} {SciPost Phys.}\ }\textbf {\bibinfo {volume}
  {2}},\ \bibinfo {pages} {003} (\bibinfo {year} {2017})}\BibitemShut {NoStop}%
\bibitem [{\citenamefont {Weinberg}\ and\ \citenamefont
  {Bukov}(2019)}]{quspin2019}%
  \BibitemOpen
  \bibfield  {author} {\bibinfo {author} {\bibfnamefont {P.}~\bibnamefont
  {Weinberg}}\ and\ \bibinfo {author} {\bibfnamefont {M.}~\bibnamefont
  {Bukov}},\ }\href {https://doi.org/10.21468/SciPostPhys.7.2.020} {\bibfield
  {journal} {\bibinfo  {journal} {SciPost Phys.}\ }\textbf {\bibinfo {volume}
  {7}},\ \bibinfo {pages} {20} (\bibinfo {year} {2019})}\BibitemShut {NoStop}%
\bibitem [{\citenamefont {Patel}\ and\ \citenamefont
  {Trivedi}(2019)}]{Nandini_pnas}%
  \BibitemOpen
  \bibfield  {author} {\bibinfo {author} {\bibfnamefont {N.~D.}\ \bibnamefont
  {Patel}}\ and\ \bibinfo {author} {\bibfnamefont {N.}~\bibnamefont
  {Trivedi}},\ }\href@noop {} {\bibfield  {journal} {\bibinfo  {journal}
  {Proceedings of the National Academy of Sciences}\ }\textbf {\bibinfo
  {volume} {116}},\ \bibinfo {pages} {12199} (\bibinfo {year}
  {2019})}\BibitemShut {NoStop}%
\end{thebibliography}%

\end{document}